\documentclass[a4paper, 12pt]{article}

\usepackage{feynmf}
\usepackage{graphicx}
\usepackage{url}
\usepackage{amsbsy,amssymb} 
\usepackage{amsmath}
\usepackage{abstract}
\usepackage{verbatim}
\usepackage{hyperref}
\usepackage{color}
 
\begin{document}

{\LARGE\centering{\bf{The adaptive Monte Carlo toolbox for phase space integration and generation}}}

\begin{center}
\sf{R. A. Kycia$^{a,b,}$\footnote{kycia.radoslaw@gmail.com}, J. Turnau, J. J. Chwastowski$^{c,}$\footnote{Janusz.Chwastowski@ifj.edu.pl}, R. Staszewski$^{c,}$\footnote{Rafal.Staszewski@ifj.edu.pl}, M.~Trzebi\'{n}ski$^{c,}$\footnote{Maciej.Trzebinski@ifj.edu.pl}}
\end{center}

\medskip
\small{
\begin{center}
$^{a~}$Tadeusz Ko\'{s}ciuszko Cracow University of Technology, \\ Warszawska 24, 31-155 Krak\'{o}w, Poland. \\
$^{b~}$Faculty of Science, Masaryk University, \\
Kotl\'{a}\v{r}sk\'{a} 2, 602 00 Brno, Czech Republic. \\
$^{c~}$Institute of Nuclear Physics  Polish Academy of Sciences, \\ Radzikowskiego 152, 31-342 Krak\'{o}w, Poland.
\end{center}
}
\bigskip

\begin{abstract}
\noindent
An implementation of the Monte Carlo (MC) phase space generators coupled with adaptive MC integration/simulation program FOAM is presented. The first program is a modification of the classic phase space generator GENBOD interfaced with the adaptive sampling
integrator/generator FOAM. On top of this tool the algorithm suitable for generation of the phase space for an reaction with two leading particles is presented (double-peripheral
process with central production of particles). At the same time it serves as an
instructive example of construction of a self-adaptive phase space generator/integrator with a modular structure for specialized 
particle physics calculations.
\end{abstract}

\setcounter{footnote}{0}

\section{Introduction}

The need for an efficient phase space generation of multiparticle final states resulted in a large number of programs. These tools offers various degree of generality of the application that improve the integration of the differential cross sections. In addition, they allow for efficient generation of events with unit weight. Such feature is needed especially when the very time consuming detector simulation is involved. These programs employ
 the sampling methods which aim at the minimisation of the integrand variance.
One example of efficient strategy is described in \cite{EWas}. The $n$-body phase space  parametrised by $3n-4$ independent variables is divided  into the manageable 
subsets (modules) to be handled by techniques which reduce the variance of the result 
(e.g. importance sampling~\cite{PeneKrzywickiCPS} or~\cite{Jadach_cylindrical} and the references therein, or the adaptive integration like 
VEGAS~\cite{VEGAS} or FOAM~\cite{Jadach-FOAM}). This strategy described in~\cite{EWas} was implemented as the parton level Monte Carlo Event 
Generator AcerMC~\cite{PeneKrzywickiCPS} with interfaces to PYTHIA 6.4~\cite{pythia6.4} and other hadronisation programs. 
Program described in this paper follows in principle the same strategy but its construction assumes the non-perturbative, e.g. Regge, hadron level 
amplitudes. Moreover, it is designed to efficiently generate the exclusive final states produced in the diffraction processes at very high energy as measured for 
example at RHIC or at the LHC.
As an example the algorithm presented in this paper as GenExLight has been applied to the reaction
\begin{equation}
 \begin{array}{c}
  p+p \rightarrow 2p + 2\pi^{+}+2\pi^{-},
 \end{array}
\end{equation}
the continuum production of the pion pairs for which one sets restrictive bounds on the transverse momenta of the final state protons -- the leading particles. For a thorough physics discussion of the above reaction see~\cite{4pi}. \\
The matrix elements for reactions described the Regge model are strongly localised in a small volume of the phase space. Therefore the generation applying the adaptive scanning of the integrand function gives one of the best, most general and flexible approach to solve such problems. To the class of the most general adaptive MC integrators belong VEGAS ~\cite{VEGAS} and FOAM ~\cite{Jadach-FOAM}. In our toolbox FOAM was selected as it is easily available within the ROOT~\cite{ROOT_site} library. However, the generators presented here can be easily adapted to use other MC integrators such as e.g. VEGAS.

The paper is organized as follows. Section \ref{Sec:PhaseSpaceGenerationAlgorithm} presents the augmented algorithm of the phase space generation
 based on the Raubold-Lynch spherical 
decay algorithm~\cite{James} and interfaced with the external program 
FOAM for adaptive Monte-Carlo integration. In section \ref{TDecay-test} a test of the considered tool efficiency is presented.
In section \ref{GenExLight} an example of the specialized self-adapting phase 
space generator (employing TDecay as a basic module, which is also described there) is presented. This code designed for the efficient generation of the double-peripheral processes with two leading particles. A test of its efficiency is presented in section \ref{GenExLight-test}. The kinematical formulae derivations and the program technical details are described
in the Appendices \ref{first appendix}, \ref{second appendix}, \ref{third appendix}. 
One should note that the methods presented below are incorporated within the GenEx (the Generator for Exclusive processes) MC code~\cite{GenEx}. However, in the present publication, they are described as the stand-alone tools, each with its own area of applications and which can be used to construct minimalistic and case-dedicated MC generators.

\section{TDecay -- spherical phase space generation algorithm}
\label{Sec:PhaseSpaceGenerationAlgorithm}
In this section  modifications of the original Raubold-Lynch algorithm~\cite{James} implemented in ROOT ~\cite{ROOT_site} as the TGenPhaseSpace class will be presented. This modification will be called further TDeacy. The changes allow interfacing the algorithm with external adaptive Monte Carlo simulators e.g FOAM. In addition,  instead of the relative weight (i.e. probability of the particular phase space configuration of final particles) we provide absolute phase space weight according to Pilkuhn convention ~\cite{Pilkuhn} (compare with \cite{Hagedorn}).

First, let us consider the integral of some function $f (P;p_{1},\ldots,p_{N})$ over the phase space \footnote{Note that in the whole paper the integration is written as ``an operator'' that ``acts'' on integrated function. It means that instead of $\int f(x) dx$ we write $\int dx f(x)$.}
\begin{equation}
 \int dLips (s; p_{1},\ldots, p_{N}) f (P;p_{1},\ldots,p_{N}) ,
 \label{GeneralPhaseSpaceIntegral}
\end{equation}
 where the sum $P$ of the four-momenta of $N$ particles $\{p_{i}\}_{i=1}^{N}$ is conserved,
 the Lorentz invariant phase space measure is given by
\begin{equation}
 dLips (s; p_{1},\ldots, p_{N}) =  (2\pi)^{4} \delta^{4}\left (P-\sum_{i=1}^{N}p_{i}\right)dLips (p_{1},\ldots,p_{N}),
\end{equation}
where $s=P^{2}$ and, finally,
\begin{equation}
 dLips (p_{1},\ldots,p_{N}) = \prod_{i=1}^{N}\frac{d^{3}p_{i}}{ (2\pi)^{3}2E_{i}}.
\end{equation}
With substitution 
\begin{equation}
 f (P;p_{1},\ldots,p_{N}) = \frac{|\mathcal{M}|^{2}}{2[\lambda(s,m_a^2,m_b^2)]^{1/2}},
\end{equation}
where $\mathcal{M}$ denotes the matrix element for process $p_a+p_b \rightarrow 
p_1+p_2+...+p_N$, $P=p_a+p_b$ and $\lambda(s,m_a^2,m_b^2)=[s-(m_a+m_b)^2][s-(m_a-m_b)^2]$,
the integral (\ref{GeneralPhaseSpaceIntegral}) represents its cross section $\sigma(s)$. The substitution  
\begin{equation}
 f (P;p_{1},\ldots,p_{N}) = \frac{|\mathcal{M}|^{2}}{2M},
\end{equation}
where $M=\sqrt{P^2}$ and $\mathcal{M}$ denotes the matrix element for the decay 
 $P \rightarrow p_1+p_2+...+p_N$, the integral in Eq. (\ref{GeneralPhaseSpaceIntegral}) represents its total width $\Gamma$.   \par
\begin{figure}[h]
\centering
\includegraphics[height=0.2\textheight, width = 0.5\columnwidth]{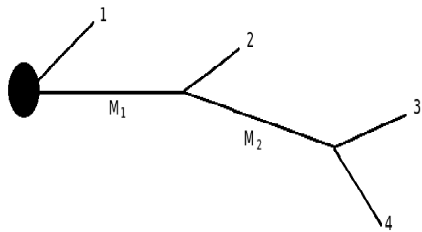}
\caption{Decay of a central mass object into four final particles. For $n>2$ final state particles there are $n-2$ intermediate particles which masses 
 are ordered $M_{1} > M_{2} > \ldots$.}
\label{Fig:Decay}
\end{figure}

The original Raubold-Lynch idea is sketched in Fig. \ref{Fig:Decay}. For a given set 
of the final particles and the decaying initial state object (marked as a black blob in Fig. \ref{Fig:Decay}) of specified four-momentum, one should generate  sequential binary decays of intermediate particles 
until all final state particles are created (see Fig. \ref{Fig:Decay}).
Each step produces a factor derived from the recurrence relation for phase space 
~\cite{Pilkuhn}:
\begin{multline}
 dLips (s;p_{1},\ldots,p_m,p_{m+1},\ldots,p_{N}) =\\dLips (s;p_{1},\ldots,p_{m},P_{r})\frac{1}{2\pi}dLips (s_{r};p_{m+1},\ldots,p_{N}),
 \label{factorisation}
\end{multline}
where the intermediate particle has four-momenta $P_{r} =\sum_{i=m+1}^{N} p_{i}$ and $s_{r} = P_{r}^{2}$. Each of $N-1$ binary decays requires two random  variables to generate the angle $\phi$ and $\cos (\theta)$ that fix the direction of the momentum vector of decaying particles. In addition $N-2$ random variables are needed to generate the intermediate masses $M_{1} > M_{2} > \ldots$, which have to be ordered to fulfil the energy conservation constraints. This is achieved using the following formulae:
\begin{equation}
M_k=\sum_{i=1}^{k}M_{i} + \left(\sqrt{s} - \sum_{i=1}^{N} M_{i} \right)x_k;\quad k=2,3,\ldots,N-1
\label{Eq:order}
\end{equation}
where $0 \leq x_{k-1}\leq x_k \leq 1$ for k=3,...N-1. Thus TDecay requires on input the 
set of $2(N-1)+ N-2= 3N-4$ random numbers (transferred from FOAM) and 
$N-2$ of them are required to be in the ascending order. The random set which is not properly ordered can be either rejected or sorted in ascending order. In the latter case $(N-2)!$ permutations of the sorted random variables correspond to $(N-2)!$ points of the FOAM probabilistic space which are mapped to single point in the physical phase space. It is therefore like the original FOAM probabilistic space was 
divided into $(N-2)!$ pieces and every one of them corresponded to the same physical phase space. Therefore, when integrating 
over such 'multiplied' FOAM probabilistic space one has to divide the integrand 
by the factor $(N-2)!$ to avoid multiple counting 
\cite{PeneKrzywickiCPS}. In the former case, when not properly ordered random sets are rejected, only $1/(N-2)!$ of the whole FOAM probabilistic space is used. Then, the efficiency of the exploration is strongly decreased, especially when the integrand has very small support in the phase space. For this reasons TDecay sorts $n-2$ variables.

The algorithm of TDecay is realized in six steps:
\begin{enumerate}
 \item {Input: rand[$3N-4$] - set of $3N-4$ random numbers; $N$ - the numbers of particles to generate; m[$N$] - array of masses of final particles; $P$ - the  decaying particle four-momenta;}
 \item {Check if the decay is energetically allowed: $\sqrt{P^{2}} > \sum_{i=1}^{N}m[i]$;}
 \item {Sort in ascending order the first $N-2$ random numbers; the additional weight factor is $\frac{1}{ (N-2)!}$;}
 \item {Generate $N-2$ intermediate masses using sorted random numbers and Eq. (\ref{Eq:order});}
 \item {For all final particles:}
 \begin{enumerate}
  \item {Calculate $|\vec{p}_{i}|$ from the two body decay of the $(i-1)$-th intermediate object to $i$-th particle and $i$-th intermediate object;}
  \item {Generate $cos (\theta)$ and $\phi$ angle using two random number and rotate $\vec{p}_{i}$ from the OZ axis into a given direction given by these variables;}
  \item {Perform a Lorentz boost from the centre of mass frame of $i-1$ intermediate object into rest frame of the whole event;}
  \item {Update the weight for the binary decay;}
 \end{enumerate}
 \item {For all final particles perform boost from the centre of mass frame of $P$ to the rest frame;}
\end{enumerate}
The above described algorithm represents only a minor modification of the TGenPhaseSpace algorithm from ROOT library. Its implementation requires detailed description which is provided in Appendix \ref{first appendix}.

\section{TDecay -- the efficiency test}\label{TDecay-test}
The program using TDecay was tested for the generation efficiency of the unweighted events (i.e. events with unit weight) measured by the time $t_{gen}$ needed to generate one such event. The performance of TDecay is compared to that of the TGenPhaseSpace, the non-adaptive implementation of the Raubold-Lynch algorithm, using the standard rejection sampling for generation of the unweighted events. Both programs perform their task in two steps. In the case of TGenPhaseSpace the maximal value of the integrand distribution $WtMax$ is searched for in the exploration step, using random sampling points in the whole domain. The next step is 
the rejection sampling  - an event is rejected if its $weight$ fulfils  
\begin{equation}
weight < randRej \times maxWt,
\label{Eq.RejectionSmpling}
\end{equation}
where $randRej \in [0;1]$ is the random number generated from the uniform distribution on the unit interval. Otherwise the event is accepted with $weight=1$. The efficiency of the random sampling increases with diminishing weight variance.

In the case of TDecay the FOAM subdivides the domain of the integrand into $ncell$ cells (see Appendix \ref{first appendix}) in such a way that the weight variance within cell is minimised during the exploration step. The rejection sampling performed cell by cell in the generation step is then more effective and with enough cells it may reach almost $100 \%$ efficiency. 

When comparing TDecay with TGenPhaseSpace it is important to set the exploration step parameters in such a way that $WtMax$ (global and by cell) is sufficiently well determined. The event generation time, $t_{gen}$, does not include the time needed to perform
the exploration step. For the comparative TDecay-TGenPhaseSpace test the number of events generated in the TGenPhaseSpace exploration step was equal to $n_{cell} \times n_{sample} = 10^8$ where $n_{cell}=10^4$ and $n_{sampl}=10^4$ are FOAM settings for TDecay run.
The number of generated events was set to $10^5$ for both programs, $|\mathcal{M}|^2=1$, 
$\sqrt(s)=200~\rm{GeV}$. Calculation has been performed for 4, 5, and 6 final state particles ($\pi$ mesons). Results are presented  
 in the Tab. \ref{Tab:FoamRejectionSamplingComparison}. 
\begin{table}
\begin{center}
\begin{tabular}{ | c | c | c | c | c | c | }
    \hline

    N & $t_{genA}[\rm{ms}]$ & $t_{genNA}[\rm{ms}]$ & $t_{genNA}/t_{genA}$ & $\sigma_A $& $\sigma_{NA}$ \\ \hline
     4 &5.3  & 21.6 & 0.25 & $8.02$ & $8.1$ \\ \hline
     5 &5.7 & 5.5 & 1.04 & $1410$ & $1410$ \\ \hline
     6 &4.0 & 4.0 & 1.00 & $46080 $ & $46060$ \\ \hline
  \end{tabular}
  \end{center}
  \caption{Comparison of times of generation of events with unit weight and values of integral  using adaptive (A) TDecay program with FOAM setting $n_{cell}=10^4$;$n_{sample}=10^4$  and non-adaptive (NA) TGenPhaseSpace with $10^8$ events in the exploration step. 
In both cases the matrix element $\mathcal{M}$ is set to 1 and $\sqrt{s}= 200 ~\rm{GeV}$. 
N is the number of generated final state particles,$t_{...}$ with subscripts genA and genNA for adaptive (A) and non-adaptive (NA)
sampling is the generation time of single event. 
 The errors for adaptive ($\sigma_A$) and non-adaptive ($\sigma_{NA})$ integral 
(in arbitrary units) are 
indicated by the last significant digit. Tests were performed on Intel Xeon E5-2680 v3 @ 2.5 GHz.}
  \label{Tab:FoamRejectionSamplingComparison}
\end{table}  

We can see that for the unit matrix element the adaptive sampling provided by FOAM offers some advantage for N=4 and almost identical generation time, $t_{gen}$, for $N=5,6$ due to insufficient number of cells in the exploration step. The advantage becomes evident when the phase space is restricted by the matrix element or for much larger $n_{cell} \times n_{sampl}$ FOAM setting. This is illustrated in the next comparative test, in which the matrix element represents the Gaussian distribution of particles with the transverse momenta $p_{ti}$ with respect to OZ axis:
\begin{equation}
  \mathcal{M}(P;p_1,p_2,...p_N)=\prod_{i=1}^{N} \exp\left ( \frac{-p_{ti}^{2}}{2\rho^{2}}\right).
 \label{MGaussInPt}
\end{equation}
We use the same program settings as above and change both the number of final particles and the departure from spherical symmetry of the matrix element, regulated by the parameter $\rho$. The advantage of the adaptive sampling increases with number of particles and depending on the $\rho$-value the improvement can reach many orders of magnitude in generation time. The results of the test are presented in the Tab. \ref{Tab:GaussianSphericalDecay}. For $N=5,6$ and $\rho \leq 5 $ TGenPhaseSpace can not produce $10^5$ events within 3 days and calculations were abandoned.
\begin{table}[h]
\begin{tabular}{ | c | c | c | c | c | c | c |}
    \hline
     N & $\rho$ [GeV] &  $t_{genA}[\rm{ms}]$ &  $t_{genNA}[\rm{ms}]$ & $t_{genA}/t_{genNA}$ & $\sigma$ A & $\sigma$ NA \\ \hline
     4 &   5    &  1.56 & 1231.3  & 0.0013 & 7.25 & 7.58 \\ \hline
     4 &   10    & 0.75 & 159.03 & 0.0047 & 2.08   & 2.08  \\ \hline
     4 &   15    & 0.64 & 74.63 & 0.0080 & 7.91 & 7.91 \\ \hline
     5 &   10    & 4.03& 456.5 & 0.009 & 1.928 & 1.93 \\ \hline
     5 &   15    & 2.66& 211.54  & 0.013 & 1.363 & 1.35 \\ \hline
      6 &   10    &  7.39 & 426.25  & 0.017 & 1.442 & 1.44 \\ \hline
     6 &   15    &  4.33 & 457.30  & 0.009 & 1.796 & 1.79  \\ \hline   
  \end{tabular}
  \caption{Comparison of the generation times of events having the unit weight and the values of integral obtained using adaptive (A) 
TDecay program with FOAM settings $n_{cell}=10^4$; $n_{sample}=10^4$  and non-adaptive (NA) TGenPhaseSpace with $10^8$ events
in the exploration step. 
In both cases the matrix element $\mathcal{M}$ is set to the multi-gauss distribution (\ref{MGaussInPt}), and $\sqrt{s}= 200 ~\rm{GeV}$.  
N is the number of generated final state particles, $t_{...}$ with subscripts genA and genNA for adaptive (A) and non-adaptive (NA)
sampling is the generation time of single event. 
 The errors of adaptive ($\sigma_A$) and non-adaptive ($\sigma_{NA})$ integral 
(in arbitrary units) are 
indicated by the last significant digit. The tests were performed on Intel Xeon E5-2680 v3 @ 2.5 GHz.}
    \label{Tab:GaussianSphericalDecay}
\end{table}  

\section{GenExLight -- adaptive phase space integration/generation }\label{GenExLight}

In principle TDecay as an adaptive sampling phase space generator/integrator is an universal tool which can deal with any matrix element, provided that FOAM is set up with a sufficient number of cells and samplings per cell. However, in many situations, application of 
the strategy outlined in \cite{EWas} (see Introduction), which combines the adaptive 
sampling with the appropriate change of some integration variables, appears to be necessary for practical reasons (computation time). The aim of the variables change is to efficiently
generate these events the most restricted either by the matrix element or by the experimental cut. In the following, we describe GenExLight, the phase space generator with a 
modular structure and the adaptive integrator. It can be treated as a light version of GenEx~\cite{GenEx} that can be quickly adapted to specialized particle physics calculations.
The code we describe below is adapted for the high energy processes in which the four-momentum transfer to two final state particles, so-called leading particles, is strongly (e.g. exponentially) restricted. The central particle production in the multi-peripheral processes ~\cite{LSf0},~\cite{LS2pi},~\cite{4pi} is an example. This code, without a complicated object oriented structure, allows even beginners to modify it to the particular applications. At the end of this section we provide examples of possible modifications.

\begin{figure}[h]
\centering
 \includegraphics[height=0.3\textheight, width = 0.8\textwidth]{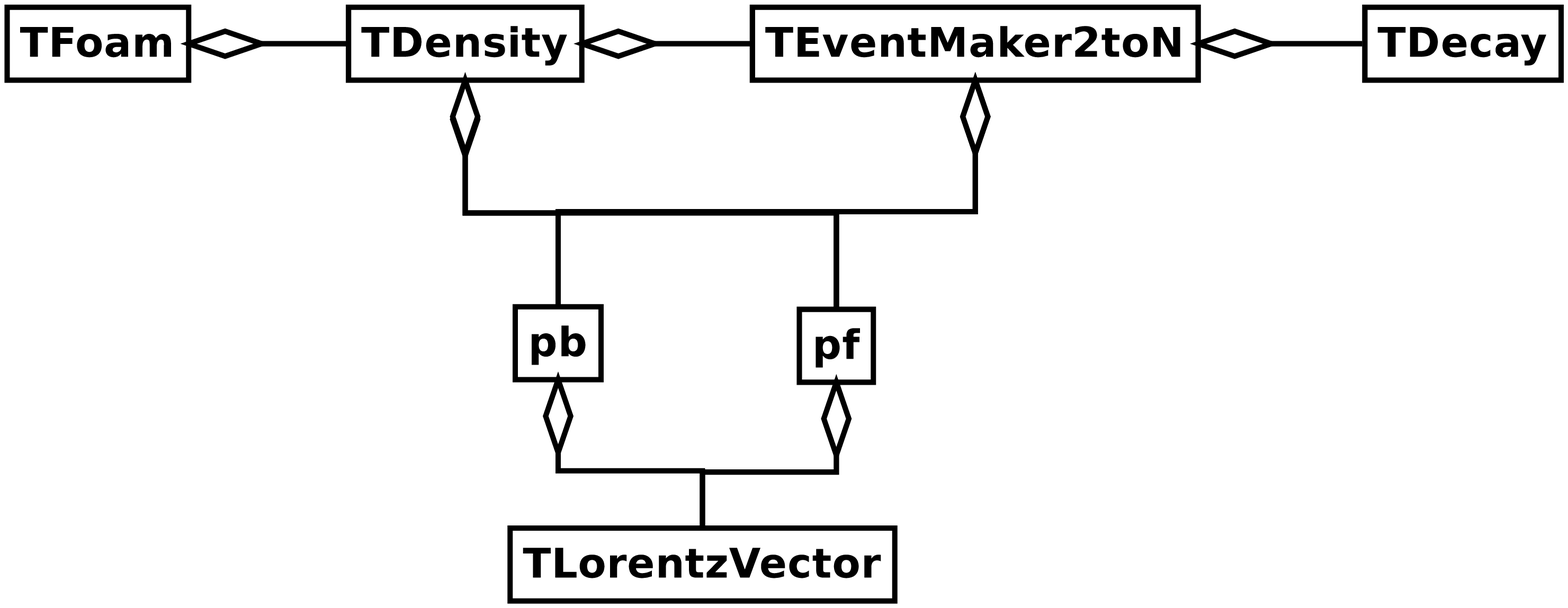}
 \caption{Class diagram of the generator.}
 \label{Fig:ClassDiagram}
\end{figure}

The structure of GenExLight is presented in Fig. \ref{Fig:ClassDiagram}. As a base it uses TDecay class described previously.
The TFoam class of adaptive Monte Carlo simulator requires TDensity class which provides integrand - a function of the four-momenta of the particles that are derived from the random numbers provided by TFoam in TEventMaker2toN. This class generates two leading particles and a central object which then is decayed by TDecay into $N-2$ remaining particles 
(see Fig. \ref{Fig:blob1}). Events are accessible in the tables $pb$ for beam particles and $pf$ for final particles that contains TLorentzVector elements.

Construction of the TEventMaker2toN class uses kinematics derived in \cite{LSf0}
for the description of the double peripheral process $pp \rightarrow ppf_0$.
The cross section formula for the reaction $p_{a}+p_{b}\rightarrow \sum_{i=1}^{N}p_{i}$ is:
\begin{equation}
 \sigma = \int   (2\pi)^{4} \delta^{ (4)}\left (p_{a}+p_{b}-\sum_{i=1}^{N}p_{i}\right) \prod_{i=1}^{N} \frac{d^{3}p_{i}}{ (2\pi)^{3}2E_{i}} \frac{|\mathcal{M}|^{2}}{2s};
 \label{crossSectionGeneral}
\end{equation}
can be transformed to the form appropriate the factorizable matrix element of the process depicted in Fig. \ref{Fig:blob1}, it is
\begin{figure}[b]
\begin{center}
\begin{fmffile}{fmftempl1}
\begin{fmfgraph*}(50,30)
   \fmfbottom{pA,pB}
   \fmftop{pA',k1,k2,k3,k4,pB'}
   \fmf{fermion,lab.side=left,
    lab=$p_a$}{pA,vA}
   \fmf{fermion,lab.side=left,
    lab=$p_{1}$}{vA,pA'}
   \fmf{fermion,lab.side=right,
    lab=$p_b$}{pB,vB}
  \fmf{fermion,lab.side=right,
   lab=$p_{2}$}{vB,pB'}
  \fmf{dbl_wiggly,lab.side=right,
   lab=$q_1$}{vA,v}
  \fmf{dbl_wiggly,lab.side=left,
   lab=$q_2$}{vB,v}
  \fmfdot{vA,vB} \fmfblob{.4w}{v}
  \fmffreeze
  \fmf{fermion,lab.side=left,
    lab=$p_3$}{v,k1}
  \fmf{fermion,lab.side=left,
    lab=$p_4$}{v,k2}
  \fmf{fermion,lab.side=left,
    lab=$p_{...}$}{v,k3}
  \fmf{fermion,lab.side=left,
    lab=$p_N$}{v,k4}
 \end{fmfgraph*}
\end{fmffile}
\caption{The illustration of the kinematics built by TEvantMaker2toN class adapted to
integration/generation of the double peripheral central production processes. 
The peripherally scattered objects $a$ and $b$ do not enter into the phase space occupied by
centrally produced particles $3,4,\ldots,N$.}
\label{Fig:blob1}
\end{center}
\end{figure}
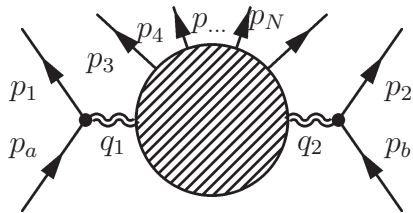
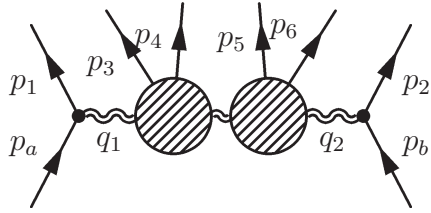
\begin{figure}[hbt]
\begin{center}
\begin{fmffile}{fmftempl2}
\begin{fmfgraph*}(50,30)
   \fmfbottom{pA,pB}
   \fmftop{pA',k1,k2,k3,k4,pB'}
   \fmf{fermion,lab.side=left,
    lab=$p_a$}{pA,vA}
   \fmf{fermion,lab.side=left,
    lab=$p_{1}$}{vA,pA'}
   \fmf{fermion,lab.side=right,
    lab=$p_b$}{pB,vB}
  \fmf{fermion,lab.side=right,
   lab=$p_{2}$}{vB,pB'}
  \fmf{dbl_wiggly,lab.side=right,
   lab=$q_1$}{vA,v1}
  \fmf{dbl_wiggly,lab.side=left,
   lab=$q_2$}{vB,v2}
  \fmfdot{vA,vB} 
  \fmfblob{.2w}{v1}
  \fmfblob{.2w}{v2}
\fmf{dbl_wiggly}{v1,v2}
  \fmffreeze
  \fmf{fermion,lab.side=left,
    lab=$p_3$}{v1,k1}
  \fmf{fermion,lab.side=left,
    lab=$p_4$}{v1,k2}
  \fmf{fermion,lab.side=left,
    lab=$p_5$}{v2,k3}
  \fmf{fermion,lab.side=left,
    lab=$p_6$}{v2,k4}
 \end{fmfgraph*}
\end{fmffile}
\caption{The illustration of possible modification of the kinematics to be built by
TEventMaker2toN class, adapted to integration/generation of the tripple 
peripheral process with the central production of two objects.}
\label{Fig:blob2}
\end{center}
\end{figure}
\begin{equation}
 \begin{array}{c}
  \sigma =  \overbrace{\int_{M^2_{min}}^{M^2_{max}} dM^{2}}^\text{A}  \overbrace{ \int   (2\pi)^{4}  \delta^{ (4)}\left (p_{a}+p_{b}- (p_{1}+p_{2}+P_{M})\right) \frac{d^{3}p_{1}}{ (2\pi)^{3}2E_{1}}\frac{d^{3}p_{2}}{ (2\pi)^{3}2E_{2}}\frac{d^{3}P_{M}}{ (2\pi)^{3}2E_{M}}}^\text{B} \\
  \underbrace{ \frac{1}{2\pi} \int  (2\pi)^{4} \delta^{ (4)}\left (P_{M}-\sum_{i=3}^{N}p_{i}\right) \prod_{i=3}^{N} \frac{d^{3}p_{i}}{ (2\pi)^{3}2E_{i}} }_\text{C} \frac{|\mathcal{M}|^{2}}{2s},
 \end{array}
 \label{crossSectionABC}
\end{equation}
where $p_1$, $p_2$ are peripheral particles and $P_{M} = \sum_{i=3}^{N}p_{i}$ is the total four-momentum of the centrally produced object which decays into $N-2$ centrally produced particles. In Eq. (\ref{crossSectionABC}) $A$ represents integration over the invariant mass square of the centrally produced object, $B$ - the integration over the 5-dimensional phase space of the $2 \rightarrow 3$ process in Fig. \ref{Fig:blob1} and $C$ - the integration
over the phase space of the decay of the central object into $N-2$ final particles. It is also assumed that the matrix element factorizes as:
\begin{equation}
\mathcal{M}(s,p_1,p_2,\ldots,p_N)=\mathcal{M}_{2\rightarrow 3}(s,p_a,p_b,p_1,p_2,p_M)\mathcal{M}_{1\rightarrow N-2}(P_M,p_3,p_4,\ldots,p_N).
\label{factorization}
\end{equation}
This assumption excludes the possibility of an interference between the identical leading and centrally produced particles as well as any matrix element factors which break factorization
in Eq. (\ref{factorization}) (e.g. some models of absorption \cite{LSabsorbtion}).  
For the double-peripheral process at sufficiently high energy, the interference
is indeed excluded, because the leading and central particles occupy 
different regions of phase space. The integral in part $B$ has to be transformed into 
the final variables appropriate for the efficient
 integration:  
\begin{equation}
 B =\int  \sum_{p'_{1z}\in\{p''_{1z}:f (p''_{z1})=0\}} \frac{ (2\pi)^{4}}{J}  \frac{|\vec{p}_{1t}|d|\vec{p}_{1t}|d\phi_{1}}{ (2\pi)^{3}2E_{1}}\frac{|\vec{p}_{2t}|d|\vec{p}_{2t}|d\phi_{2}}{ (2\pi)^{3}2E_{2}}\frac{dy_{M}}{ (2\pi)^{3}2}.
\label{integral B}
\end{equation}
where $\phi_{1}$ and $\phi_{2}$ are the polar angles, $y_{M}$ is the rapidity of the central object and 
$J=\left|\frac{p_{1z}}{E1}+\frac{p_{2z}}{E_{2}}\right|$. The summation goes over the 
solutions of the quadratic equation $f (p_{1z})=0$ where  
\begin{equation}
 f (p_{1z})=\sqrt{s}-\sqrt{p_{1t}^{2} + p_{1z}^{2}}-\sqrt{p_{2t}^{2} + p_{2z}^2 }-E_{M}.
\end{equation}
If $\sqrt{s}>> E_M$ we deal with almost two-body process, the solutions correspond to two parts of the phase space in which $p_{az}p_{1z}>0$  (and  $p_{bz}p_{2z}>0$) and $p_{az}p_{1z}<0$ (and $p_{bz}p_{2z}<0$). In the very high energy double peripheral processes the initial particles do not ``bounce'' from each other and only the first solution contributes to the integral. However, in a general case both solutions should be equally treated. This is achieved by adding an additional random variable that will select one of the  two of solutions for each generated event. The derivation of the Eq. (\ref{integral B}) is shown in Appendix \ref{second appendix}. For the technical description of GenExLight see Appendix \ref{third appendix}.

GenExLight should be considered as a particular realization of the 
strategy for a modular construction of the effective generators with an adaptive sampling using 
TDecay as a basic module. Let us reiterate the essential elements of the program structure. 
TEventMaker2toN class builds kinematics adapted to the process of interest. In our 
example such process is double peripheral central production of particles, as depicted in Fig. \ref{Fig:blob1}. The simplest modification which can be introduced is to allow for a
dissociation of the leading particles i.e. by replacing each of them by intermediate objects of variable mass which are subsequently decayed using TDecay. In order to efficiently generate a triple peripheral process with production of two central objects and two leading particles, 
Fig. \ref{Fig:blob2}, one could construct TEventMaker2toN based on kinematics of $2 \rightarrow 4$ using formulae derived in ~\cite{LS2pi}. Again, one should remember about the factorization assumption of the matrix element we have made. It excludes possibility of the interference of identical final state particles which belong to different decaying objects.
For this reason in Ref. \cite{4pi} $2 \rightarrow 3$ kinematics of  TEventMaker2toN was employed. Although at sufficiently high invariant mass of the four-pion system  $2 \rightarrow 4$ kinematics would be acceptable approximation resulting in much better
generator efficiency.

\section{GenExLight -- performance test}\label{GenExLight-test}
For the GenExLight performance test we select reaction $p_{a} (p)+p_{b} (p) \rightarrow p_{1} (p)+p_{2} (p) + p_{3} (\pi^{+})+p_{4} (\pi^{-})$ at $\sqrt{s} = \sqrt{ (p_{a}+p_{b})^{2}} = 200~\rm{GeV}$ with the unit matrix element and the following cuts: 
\begin{itemize}
 \item {$|t_{1}|=|(p_{a}-p_{1})^{2}| < t_{cut}$, $|t_{2}|=|(p_{b}-p_{2})^{2}| < t_{cut}$}
 \item {$M_{3+4} \in [0.280~\rm{GeV}; 100~\rm{GeV}]$}
 \item {$|y_{3+4}| < 10$}
\end{itemize}
where $t_{cut}$ is a parameter. The cuts restrict the four-momentum transfers  $ a \rightarrow 1 $ and $b \rightarrow 2$ and indirectly transverse momenta of particles $1$ and $2$, as well as, the variables directly generated in TEventMaker2toN. In addition, the mass and rapidity of the centrally produced object are restricted, so the resulting kinematics resembles that of the typical double peripheral process e.g. \cite{LS2pi}. We compare performance of GenExLight and TDecay in terms of exploration $t_{expl}$ and generation $t_{gen}$ times of a single event with unit weight. 
\begin{table}[t]
\begin{center}
\begin{tabular}{ | c | c | c |c | c | c | }
    \hline
    $t_{cut}$[GeV] & $\sigma_{GEL}$& $t_{explGEL}$[s] & $\sigma_{TD}$ & $t_{explTD}$[s] & $t_{genTD}/t_{genGEL}$ \\ \hline
        200    &  $4.469$ & $101.73$ & $1.399$ & $301.76$ & 0.024 \\ \hline
        300    &  $9.53$  & $90.5$   & $6.54$  & $295.04$ & 0.040\\ \hline
        400    &  $1.540 $ & $102.11$ & $1.254$ & $321.5$&  0.046 \\ \hline
        500    &  $2.109 $ & $99.41$  & $2.501$ & $320.69$ & 0.054 \\ \hline
        600    &  $2.69  $ & $91.18 $  & $3.710 $ & $306.96$& 0.07 \\ \hline
        700    &  $3.298$ & $83.93 $  & $4.99$  & $303.1 $& 0.07 \\ \hline
        800    &  $6.73 $  & $109.96 $ & $6.52 $  & $298.55 $ & 0.09\\ \hline
        900    &  $8.44 $  & $122.43 $ & $8.25 $  & $272.06$ & 0.13  \\ \hline
        1000   &  $1.02 $ & $117.91 $ & $1.01 $ & $266.41 $& 0.13 \\ \hline
  \end{tabular}
  \caption{Efficiency comparison GenExLight (GEL) and TDecay (TD) with identical FOAM parameter
  settings (see text) . $t_{expl...}$ is the exploration time and $t_{gen...}$ is the generation time of single event with unit 
weight. Results were obtained for $\mathcal{M}=1$ and $\sqrt{s}= 200 ~\rm{GeV}$, cuts and the process described in the text.
Tests were performed on Intel Xeon E5-2680 v3 @ 2.5 GHz.}
\label{Tab:ConvergenceT1}
\end{center}
\end{table}
 The FOAM exploration parameters are set to $10^{4}$ cells and $10^{4}$ samples in each cell 
and the number of generated events is $10^{5}$. The results are presented in Tab. \ref{Tab:ConvergenceT1}.

The table shows that for large values the cut $t_{cut} > 700 GeV^{2}$ i.e. when the final state sphericity is not too small, both algorithms converge and provide results differing by less than $3\%$. However, agreement within uncertainty indicated by FOAM is reached only for $t_{cut} > 1000 GeV^{2}$. For smaller values of $t_{cut}$, i.e. more asymmetric final state, TDecay program performs poorly and to get a meaningful result the FOAM parameters have to be set to much more cells and samplings/cell. For $t_{tcut}=200 GeV^{2}$ with $n_{cell}=10^5, n_{sampl}=10^5$ the TDecay-GenExLight convergence is achieved within indicated statistical uncertainty (see Tab. \ref{Tab:ConvergenceT2}).
\begin{table}[h!]
\begin{center}
\begin{tabular}{ | c || c | c || c | c || c | }
    \hline
    $n_{cell} \times n_{sampl}$ & $\sigma_{GEL}$& $t_{explGEL} $ [min]  & $\sigma_{TD}$& $t_{explTD}$[min]& $t_{genTD}/t_{genGEL}$ \\ \hline
    $10^{4}\times 10^{4}$ &  4.469 & 1.7 & 1.399 & 5.02  & 0.024  \\ \hline \hline
    $10^{5}\times 10^{5}$ &  4.468 & 94.5 & 4.30 & 349.8 & 0.04 \\ \hline 
  \end{tabular}
  \caption{Efficiency comparison GenExLight (GEL) and TDecay (TD) with two different  FOAM parameter $n_{cell}$ and $n_{sampl}$
  settings. $t_{expl...}$ is the exploration time and $t_{gen...}$ is the generation time of single event with unit 
  weight. Results were obtained for $\mathcal{M}=1$, $\sqrt{s}= 200 ~\rm{GeV}$, cuts and the process described in the text.
Tests were performed on Intel Xeon E5-2680 v3 @ 2.5 GHz.}
    \label{Tab:ConvergenceT2}
\end{center}
\end{table}  
The exploration time for TDecay increased by a factor $70$ while  the generation time of a single event with an unit weight is $25$ times longer than in the GenExLight case. GenExLight performs perfectly with $n_{cell}=10^4, n_{sampl}=10^4$ FOAM settings even for the realistic experimental cut $t_{cut} < 1$. It should be noted that the uncertainties indicated by FOAM are not reliable when the number of cells and samplings/cell are too small. In practice, one should  use the generator with a few set-ups of exploration increasing the number of cell until results get stable.

\section{Conclusions}\label{Conclusions}
In this paper a simple and robust toolbox for phenomenological studies of exclusive
hadronic reactions was discussed. Its basic element, TDecay class, the module
 for the specialized self adapting event generators, was described. TDecay is a modification of the classic phase space generator GENBOD (implemented as TGenPhaseSpace in ROOT) interfaced with the adaptive sampling integrator/generator FOAM. A practical example of its usage in construction of GenExLight, the program designed for integration/generation of the central production in double peripheral process, was presented, together with comparative performance tests. These tests show not only the power of adaptive sampling, but also its limitations. It appears, that for purely practical reasons (the computation time) appropriate change of variables which enables the program to limit the number of empty cells is often a necessity.

\appendix

\section{Appendix: Technical description of TDecay class}\label{first appendix}

TDeacy generates $n$-body decay events $ P \rightarrow p_1,p_2,...,p_n$ and is 
interfaced with FOAM, the adaptive sampling Monte-Carlo generator/integrator. It is also an utility class which can be combined with dedicated sampling algorithms, as described in Section \ref{GenExLight}. The program can be downloaded from \cite{Repository} in the directory TDecayTFoam. The directory contains the TDecay class (header and implementation files), the main file as well as a Makefile.

The TDecay interface consists of three functions.
\begin{itemize}
 \item {Function that specifies the decay configuration is
\begin{verbatim}
  Bool_t SetDecay (TLorentzVector &P, Int_t nt, const Double_t *mass)
\end{verbatim}
where $P$ is four-momentum of the decaying particle, $nt$, $mass$ - 
number of final particles and their masses in an array.}
\item {Function 
\begin{verbatim}
Double_t  Generate (std::queue<double> &rnd)
\end{verbatim}
derives four-momenta and the phase space weight (see $B$ in Eq. (\ref{crossSectionABC}))
from $rnd$ - standard C++ library queue of  $3\times nt - 4$  double precision 
random numbers. The weight contains also the factor $\frac{1}{2\pi}$ as it is 
indicated in Eq. (\ref{factorisation}). The four-momenta of the final particles can be retrieved by the function
\begin{verbatim}
 TLorentzVector *GetDecay (Int_t n)
\end{verbatim}
where $n$ is the index of $n$-th final particle starting from $0$.}
\end{itemize}

The basic parameters defining reaction are at the top of main file.
\begin{verbatim}
//center of mass energy
const double tecm = 200.0;

//incoming particles
const int Nip = 2; 
//outgoing particles
const int Nop = 4;

//pdg codes of particles starting from index 1
int idIn[Nip+1] = {0, 2212,2212};	     //PDGID incoming paricles
int idOut[Nop+1] = {0, 2212,2212,211,-211}; //PDGID outgoing paricles
\end{verbatim}

The TDensity class inherits from TFoamIntegrand. It is required by FOAM, defines how 
a set of random numbers is transformed into the value of integrand; for 
details of this structure see \cite{Jadach-FOAM}. The most important method 
of this class is the TDensity::Density function which uses TDecay object 
in order to generate events written in $pb$ and $pf$ tables, place experimental 
cuts and calculate integrand function - matrix element for the given reaction multiplied by phase space weight.

At the beginning of the main function there is a setup for FOAM including the number of events 
\begin{verbatim}
long NevTot   =     100000;   // Total MC statistics
\end{verbatim}
and the exactness of the exploration procedure
\begin{verbatim}
Int_t  nCells   =     10000;   // Number of Cells
Int_t  nSampl   =     10000;   // Number of MC events per cell in build-up
\end{verbatim}

The next section defines sample histograms and ROOT tree used to collect data. After this part the main integration loop follows. First part of the loop is the generation of the event and associated weight
\begin{verbatim}
 // generate MC event
FoamX->MakeEvent ();           
FoamX->GetMCvect ( MCvect);
FoamX->GetMCwt ( MCwt );
\end{verbatim}
then the histograms and tree are filled with the generated event and then the integrand is added to the sum of weights generated in the previous rounds of the loop. At the end of the program the booking actions are taken - printing the integral, saving histograms, tree and freeing memory from the allocated objects.

The program is supplied with Makefile~\cite{GNUMAKE}, with the following commands:
\begin{itemize}
 \item {make run - compile and run the program;}
 \item {make clean - clean directory from executables;}
 \item {make cleanest - clean directory from executables, eps figures and root tree with events;}
\end{itemize}

\section{Appendix: $2 \rightarrow 3$ kinematics in TEventMaker2toN} \label{second appendix}
In this Appendix we show steps transforming the part $B$ of the cross section formula Eq. (\ref{crossSectionABC}) to Eq. (\ref{integral B}) which is kinematical basis of the TEventMaker2toN class. The calculations are made in the reaction centre of mass  ($\vec{p}_{a}+\vec{p}_{b}=0$) and the beam is aligned along OZ axis. Integration over the transverse momentum  of the central object 
$\vec{P}_{Mt}$ 
gives
\begin{equation}
 B=\int  (2\pi)^{4}  \delta (\sqrt{s}-E_{1}-E_{2}-E_{M})\delta\left (p_{1z}+p_{2z}+P_{Mz}\right) \frac{d^{3}p_{1}}{ (2\pi)^{3}2E_{1}}\frac{d^{3}p_{2}}{ (2\pi)^{3}2E_{2}}\frac{dP_{Mz}}{ (2\pi)^{3}2E_{M}},
\end{equation}
with the constraint $\vec{P}_{mt}=- (\vec{p}_{1t}+\vec{p}_{2t})$. In the next step, integration over $p_{2z}$ gives
\begin{equation}
 B=\int  (2\pi)^{4}  \delta (\sqrt{s}-E_{1}-E_{2}-E_{M}) \frac{d^{3}p_{1}}{ (2\pi)^{3}2E_{1}}\frac{d^{2}p_{2t}}{ (2\pi)^{3}2E_{2}}\frac{dP_{Mz}}{ (2\pi)^{3}2E_{M}},
\end{equation}
with the constraint $p_{1z}+p_{2z}+P_{Mz}=0$. In order to perform the last integration over $p_{1z}$ the argument of the remaining Dirac delta has to be expressed in terms of this variable. Let us introduce
\begin{equation}
 f (p_{1z})=\sqrt{s}-\sqrt{p_{1t}^{2} + p_{1z}^{2}}-\sqrt{p_{2t}^{2} + p_{2z}^{2}}-E_{M},
\end{equation}
where $p_{2z} (p_{1z})=- (p_{1z}+p_{Mz})$. Now $\delta (\sqrt{s}-E_{1}-E_{2}-E_{M}) = \delta (f (p_{z1}))$. The solution of the quadratic equation $f (p_{1z})=0$ generates two solutions discussed in Eq. (\ref{GenExLight}).

Using the standard formula, the Dirac delta can be transformed in the following way
\begin{equation}
 \delta (f (p_{z1})) = \sum_{p'_{1z}\in\{p''_{1z}:f (p''_{z1})=0\}}\frac{\delta (p_{1z}-p'_{z1})}{J},
\end{equation}
where $J=\left|\frac{p_{1z}}{E1}+\frac{p_{2z}}{E_{2}}\right|$. At this stage integration over $p_{1z}$ is easily performed and results in
\begin{equation}
 B=\int  \sum_{p'_{1z}\in\{p''_{1z}:f (p''_{z1})=0\}} \frac{ (2\pi)^{4}}{J}  \frac{d^{2}p_{1t}}{ (2\pi)^{3}2E_{1}}\frac{d^{2}p_{2t}}{ (2\pi)^{3}2E_{2}}\frac{dP_{Mz}}{ (2\pi)^{3}2E_{M}}.
\end{equation}

Last part is transformation to polar variables in $\vec{p}_{1t}$ and $\vec{p}_{2t}$ and to $P_{Mz} \rightarrow y_{M}$, where $y_{M}$ is the rapidity of the central object. That results in final form
\begin{equation}
 B =\int  \sum_{p'_{1z}\in\{p''_{1z}:f (p''_{z1})=0\}} \frac{ (2\pi)^{4}}{J}  \frac{|\vec{p}_{1t}|d|\vec{p}_{1t}|d\phi_{1}}{ (2\pi)^{3}2E_{1}}\frac{|\vec{p}_{2t}|d|\vec{p}_{2t}|d\phi_{2}}{ (2\pi)^{3}2E_{2}}\frac{dy_{M}}{ (2\pi)^{3}2},
\end{equation}
where $\phi_{1}$ and $\phi_{2}$ are polar angles.

\section{Appendix: Technical description of GenExLight program}\label{third appendix}

In this appendix the interface of the  classes in  Fig. \ref{Fig:ClassDiagram} is described.
The constructors of TDensity and TEventMaker2toN are as follows
\begin{verbatim}
TEventMaker2toN (double tecm, double p_min, double p_max, double y_min, double y_max, 
int nop, double mass_min, double mass_max, double* mass, int* idIn, int* idOut, 
TLorentzVector* pb, TLorentzVector* pf, int isol)
 
TDensity (double tecm, double p_min, double p_max, double y_min, double y_max, 
int nop, double mass_min, double mass_max, double* mass, int* idIn, int* idOut, 
TLorentzVector* pb, TLorentzVector* pf, int isol);	
\end{verbatim}
where the meaning of the variables is given in Tab. \ref{Tab:ConstructorArguments}.

  \begin{table}
\begin{center}
    \begin{tabular}{| l | p{0.8\textwidth} | }
    \hline
     Variable & Meaning \\ \hline
     $tecm$ & centre of mass energy $\sqrt{ (pb_{1}+pb_{2})^{2}}$; \\ \hline
     $p\_min$, $p\_max$  & the range of transverse momenta of leading particles $1$ and $2$; \\ \hline
     $y\_min$, $y\_max$  & the range of rapidity for the central object; \\  \hline
     $nop$  & number of outgoing particles; \\ \hline
     $mass\_min$, $mass\_max$ & range of the mass of central object; \\ \hline
     $mass$ & mass matrix for central particles; first particle has index $1$; the index $0$ is unused; \\ \hline
     $idIn$ & table of PDG identifiers for incoming particles $pb_{1}$ and $pb_{2}$; \\ \hline
     $idOut$ & table of PDG identifiers for outgoing particles $pf_{1},\ldots,pf_{N}$; \\ \hline
     $pb$ & array of four momenta of incoming particles; \\ \hline
     $pf$ & array of four momenta of outgoing particles; \\ \hline
     $isol$ & selecting solution for kinematics: 0 - nonbouncing, 1- bouncing \\ \hline
     \end{tabular}
\caption{Variables in constructors of TEventMaker2toN and TDecay.} 
     \label{Tab:ConstructorArguments}
  \end{center}
   \end{table}

Generator can be started and cleaned using GNUMake system~\cite{GNUMAKE} with the following commands:
\begin{itemize}
 \item {make run - compile and run generator;}
 \item {make clean - clean executables and intermediate files;}
 \item {make cleanests - clean executables and results of simulation;}
\end{itemize}
Generator produces the following output files:
\begin{itemize}
 \item {events.root - ROOT file with events as TLorentzVector;}
 \item {CM.eps - plot of central mass;}
 \item {etaPhi.eps - plot of $\eta$ vs $\phi$ for all particles in the events;}
 \item {rapidity.eps - plot of particles;}
 \item {histograms.root - histograms in ROOT file;}
\end{itemize}

\section*{Acknowledgments}
We would like to thanks Piotr Lebiedowicz and Antoni Szczurek for discussions. This work was supported in part by Polish National Science Center grant: UMO-2015/17/D/ST2/03530.  J. J. C. was partially supported by Polish National Science Center grant: UMO-2015/19/B/ST2/00989. The calculations presented in this article were supported in part by PLGrid and Cracow Cloud One infrastructures. R. K. was partially supported by the grant MUNI/A/1103/2016 of Masaryk University.




\end{document}